\documentstyle[12pt]{article}
\input{psfig} 
\input epsf.tex

\def\1ad{\mbox{\normalsize $^1$}}
\def\2ad{\mbox{\normalsize $^2$}}
\def\3ad{\mbox{\normalsize $^3$}}
\def\4ad{\mbox{\normalsize $^4$}}
\def\5ad{\mbox{\normalsize $^5$}}
\def\6ad{\mbox{\normalsize $^6$}}
\def\7ad{\mbox{\normalsize $^7$}}
\def\8ad{\mbox{\normalsize $^8$}}
\def\makefront{\vspace*{1cm}\begin{center}
\def\newtitleline{\\ \vskip 5pt}
{\Large\bf\titleline}\\
\vskip 1truecm
{\large\bf\authors}\\
\vskip 5truemm
\addresses
\end{center}
\vskip 1truecm
{\bf Abstract:}
\abstracttext
\vskip 1truecm}
\begin {document}
\rightline{hep-th/9801109}
\rightline{LBNL-41189, UCB-PTH-97/66}
\def\titleline{ 
String Duality and
\newtitleline
Novel Theories without Gravity\footnote{Based on
a talk given at the \lq\lq 31st International
Symposium Ahrenshoop on the Theory of Elementary
Particles," Buckow, Germany, September 2-6 1997} 
}
\def\authors{
Shamit Kachru  
}
\def\addresses{
Department of Physics \\
University of California, Berkeley \\
and\\
Lawrence Berkeley National Laboratory\\
University of California\\ 
Berkeley, CA 94720, USA\\
}
\def\abstracttext{
We describe some of the novel 6d quantum field theories
which have been discovered in studies of string duality.
The role these theories (and their 4d descendants)
may play in alleviating the vacuum degeneracy problem in 
string theory is reviewed.  The DLCQ of these field theories
is presented as one concrete way of formulating them, 
independent of string theory. 

}
\makefront
\section{Introduction}

Recent advances in string theory have led to the discovery
of many new interacting theories without gravity.  These theories
are found by taking special limits of M-theory, in which
many of the degrees of freedom decouple.  In this talk we will: 

\noindent
I. Describe some examples of these new theories.

\noindent
II. Review why it is important to fully understand these examples.

\noindent
III.  Propose a definition of these theories, in the light-cone
frame, which is manifestly independent of M or string theory.

\section{Examples}

\subsection{Theories with $(2,0)$ Supersymmetry}

The first (and simplest) examples were found by Witten \cite{sixd}, in
studying type IIB string theory on $K3$.  He considers the situation
where the $K3$ develops an $A-D-E$ singularity.  In the IIA theory,
one finds extra massless gauge bosons in these circumstances.  
These extra vectors of the (1,1) supersymmetry are required by
string-string duality, and arise from D2 branes which wrap the
collapsing 2-cycles and become massless in the singular limit.

In the IIB theory, there is a chiral (2,0) supergravity in six
dimensions.  The only massless multiplet of the (2,0) supersymmetry
(other than the gravity multiplet) 
is the tensor multiplet, which consists of 5 scalars, some chiral 
fermions, and a self-dual two form $B_{\mu\nu}$
which satisfies
\begin{equation}
dB = * dB
\label{1}
\end{equation}  

The (2,0) supergravity $\it requires$ the presence of precisely
21 tensor multiplets for anomaly freedom.  Therefore, it is hard
to envision a scenario where one finds extra massless particles
at the singular point in moduli space.  However, further compactification
on an $S^1$ yields a theory related to the IIA theory by T-duality,
so one must find (after $S^1$ compactification) gauge bosons of
the $A-D-E$ gauge group.  What is their IIB origin?

Further compactify the IIA and IIB theories on circles with radii 
$R_{A,B}$.  Then T-duality relates the theories
with $R_{A} = {1\over R_{B}}$ (we are temporarily 
setting the string scale $\alpha^\prime$ to one for simplicity).  
The relation between 
the six-dimensional string couplings
$\lambda_{A,B}$ is 
\begin{equation}
{1 \over \lambda_{A}} = {R_{B}\over \lambda_{B}}
\label{2}
\end{equation}
If we consider a point in IIA moduli space a distance $\epsilon$
from the singular point, then there are W-bosons coming from wrapped
D2 branes whose masses go like
\begin{equation}
M_{W} = {\epsilon \over \lambda_{A}}
\label{3}
\end{equation}
So in type IIB, the mass is
\begin{equation}
M_{W} = {\epsilon R_B \over \lambda_{B}}
\label{4}
\end{equation}
This looks like the mass of a $\it string$ wrapped around the
$S^1$ in the IIB theory!  But this string is not the critical
type IIB string; from equation (4) it must have a tension
\begin{equation}
T = {\epsilon \over \lambda_{B}}
\label{5}
\end{equation}
Of course, this string comes from a D3 brane wrapped around
a collapsing sphere in the $K3$ of area $\simeq \epsilon$.

For very small $\epsilon$, $T << {1\over \alpha^\prime}$.  So we get
an $A-D-E$ series of quantum theories in six dimensions
which contain light string solitons.  Because the noncritical 
strings 
are very light compared to the fundamental
string scale, one can decouple
gravity.  Then, it is believed that one is left with an
interacting quantum field theory in six dimensions.
As $\epsilon \rightarrow 0$, one approaches a nontrivial
fixed point of the renormalization group. 

In six-dimensions, strings are dual to strings.  The particular
light strings in question are self-dual (the $H=dB$ they produce
is self-dual as in equation (1)), 
so the \lq\lq coupling" of these quantum theories is fixed and
of order one.  In other words, there is no coupling constant which
can serve as an expansion parameter.

By using ALE spaces instead of $K3$, one can find such interacting
theories for each $A_k$ or $D_k$ singularity.
For the $A_k$ theories, there is another simple description due
to Strominger and Townsend \cite{st,town}.  For instance,
consider two parallel
M5 branes
in eleven-dimensional flat spacetime.  There are membranes which
can end on the M5 branes, 
yielding a noncritical string on the
fivebrane worldvolume with tension proportional to
the separation.  

\vskip .25cm
$$
\vbox{
{\centerline{\epsfxsize=3in \epsfbox{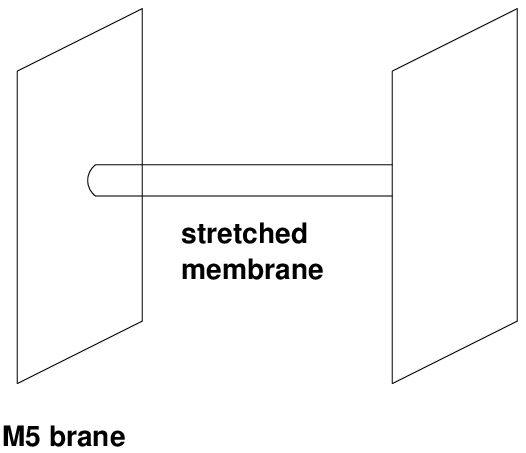}}}
{\centerline{ Figure 1:
A membrane stretching between two M5 branes.}}
}
$$
\vskip .25cm

Each fivebrane has a tensor multiplet
on its worldvolume (the five scalar components parametrize
the transverse position of the fivebrane in eleven dimensions).
If we denote the two five-tuples of scalars by $\vec \phi_{1,2}$, then 
the VEVs $\langle \vec \phi_{1,2}\rangle$ label different
vacua of an effective six-dimensional theory. 
When $\vec \phi_{1} \rightarrow
\vec \phi_{2}$, the noncritical strings become tensionless
and we find another description of the $A_1$ fixed point
above.   

More precisely, if we say the fivebranes are
separated by a distance $L$, the limit one wishes it to take is
\begin{equation}
M_{pl} \rightarrow \infty,~~L  \rightarrow 0,
~~T = LM_{pl}^{3}~fixed
\label{6}
\end{equation}
In this limit, gravity decouples but the noncritical strings
stay light, yielding an interacting theory without gravity.
The obvious generalization with $k$ fivebranes yields the
$A_{k-1}$ (2,0) fixed point, with moduli space
\begin{equation}
{\cal M}_{k} = {R^{5k} \over S_{k}}
\label{7}
\end{equation}
given by the positions of the parallel fivebranes, mod permutations.
At generic points on ${\cal M}_{k}$, the low energy theory has
$k$ tensor multiplets.

\subsection{Theories with $(1,0)$ Supersymmetry}

New interacting 6d theories with (1,0) supersymmetry have also
been discovered \cite{ganhan,sw}.  
Perhaps the simplest example is the following.
Consider Horava and Witten's description of the 
$E_8 \times E_8$ heterotic string
as M-theory on $S^1/Z_2$ \cite{horwit} . 
The length of the interval is related to the heterotic $g_{s}$,
while $E_8$ gauge fields live on each of the two \lq\lq end of
the world" ninebranes.

We can consider a fivebrane at some
point on the interval.  Its position in the $S^1/Z_2$ is parametrized
by the real
scalar $\phi$ in a (1,0) tensor multiplet, while its other transverse
positions are scalars in a (1,0) hypermultiplet.
Since it is a scalar in six dimensions, $\phi$ naturally has
dimension two; we will say the two $E_8$ walls
are located at $\phi = 0$ and $\phi = {1\over \alpha^\prime}$.
Then, one 
has noncritical strings on the fivebrane world volume with
tensions
\begin{equation}
T_{1} = \phi,~~T_{2} = ({1\over \alpha^\prime}-\phi)
\label{8}
\end{equation}
coming from membranes with one end on the fivebrane and one end on
the ninebranes.  

\vskip .25cm
$$
\vbox{
{\centerline{\epsfxsize=3in \epsfbox{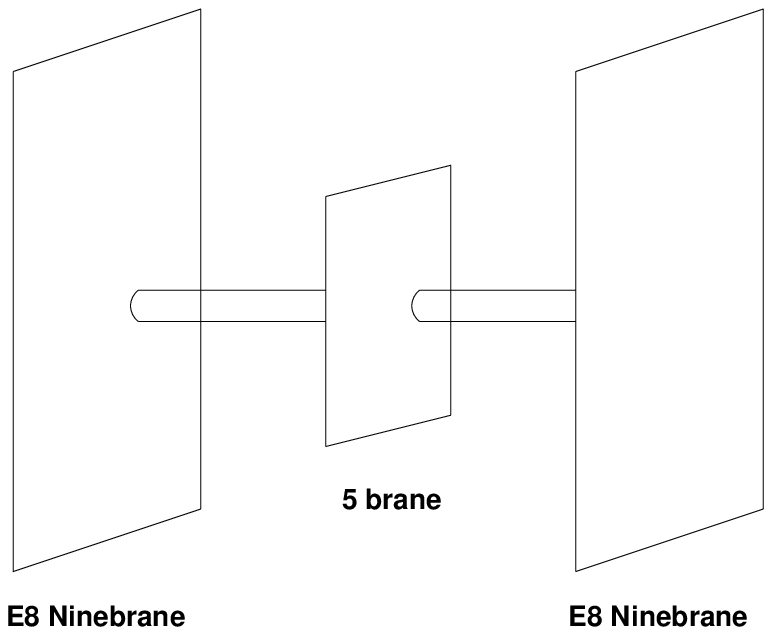}}}
{\centerline{ Figure 2:
A fivebrane between two ninebranes and two 5-9 membranes.}}
}
$$
\vskip .25cm
                      
As $\phi \rightarrow 0$ or $\phi \rightarrow {1\over\alpha^\prime}$,
one again finds that the lightest degrees of freedom in the theory
are solitonic self-dual noncritical strings.  The theory on the
fivebrane needs to also have an $E_8$ global symmetry, to couple 
consistently to the $E_8$ gauge fields on the end of the world.
Therefore, one concludes that when the fivebrane hits the ninebrane,
one finds a nontrivial (1,0) supersymmetric RG fixed point, with
$E_8$ global symmetry.

For both the (2,0) $A_k$ theories and the $E_8$ theory in 6d,
there is no known UV free Lagrangian which flows, in the IR,
to the fixed point of interest.  Therefore, it is of intrinsic
interest to find a definition of these quantum field theories
which is independent of string or M-theory.  We will propose
such a definition in \S4.
Before doing that, it seems proper to provide some motivations
for the study of these theories.

\section{Why are these theories of interest?}

There are at least three motivations for studying these theories:

\noindent
a) These are the first examples of nontrivial fixed point quantum
field theories above four dimensions.  For instance, if one
does a very naive analysis of  
gauge field theory in $d$ dimensions
\begin{equation}
L ~=~\int d^{d}x ~{1\over g^2} F_{\mu\nu}^{2} + \cdots
\label{9}
\end{equation}
one finds that $[g^2] = 4-d$ (in mass units), so the theory is
$\it infrared~free$ for $d>4$.   
Hence, the theories of \S2 are of intrinsic interest as a new
class of interacting quantum field theories.

\noindent
b) These theories play a crucial role in the study of M(atrix)
theory compactifications.  In M(atrix) theory, one starts with
the maximally supersymmetric D0 brane quantum mechanics, with
$N$ zero branes giving the DLCQ in a sector with light-like momentum
$N$ and the $N\rightarrow \infty$ limit yielding the 11d uncompactified
theory \cite{bfss,susskind}.  
To study compactifications on a transverse $T^n$, one then
T-dualizes the $U(N)$ D0 brane quantum mechanics to obtain a description
of M-theory on $T^n$ as the $n+1$ dimensional $U(N)$ 
Super Yang-Mills theory
compactified on the dual torus, $\tilde T^n$.

An obvious problem with this approach is that for $n>3$, the 
Super Yang-Mills is ill-defined at short distances (it is not
renormalizable).  
Let us consider the first such case: M-theory on $T^4$.  The
U-duality group in this case is $SL(5,Z)$.  This suggests
that perhaps the M(atrix) definition involves some 5+1 dimensional
QFT compactified on a $\tilde T^5$, 
geometrizing the $SL(5,Z)$ U-duality
group as the modular group of the torus.
The unique candidate which is well-defined (and has the correct
supersymmetry) is the $A_N$ (2,0) quantum field theory of \S2
\cite{rozali,brs}.
But, how does this prescription relate to our expectation that
the theory should be 4+1 $U(N)$ SYM?

The 5d SYM theory has a conserved $U(1)$ current
\begin{equation}
j = * (F\wedge F)
\label{10}
\end{equation}
We can identify $j$ with the Kaluza-Klein $U(1)$ symmetry of the (2,0)
theory compactified on an $S^1$ of radius $\tilde L_{5}$, if
we say
\begin{equation}
2\pi \tilde L_{5} ~=~{g^{2}\over 2\pi}
\label{11}
\end{equation}
The 5d theory has particles which are 4d instantons, and whose
action is given by $4\pi^2 {n\over g^{2}}$ for the \lq\lq n-instanton"
particle.  These in turn can be interpreted as Kaluza-Klein modes coming from
the (2,0) theory with a momentum $p_{5} = {n\over \tilde L_5}$
around the hidden \lq\lq extra" circle which promotes $\tilde T^4$
to $\tilde T^5$.  
In this way, one ends up with the prescription that M-theory on
$T^4$ is defined by the (2,0) theory on $\tilde T^5$ \cite{rozali,brs}.  
This
makes the SL(5,Z) U-duality manifest.

\noindent
c) These novel interacting theories play a crucial role in
the unification of M-theory vacua.  Consider, for instance,
the heterotic $E_8 \times E_8$ theory compactified on $K3$. 
There is a Bianchi identity for the three-form field strength
$H$ which looks like 
\begin{equation}
dH = Tr(R\wedge R) - Tr(F\wedge F)
\label{12}
\end{equation} 
where $R$ is the curvature and $F$ is the Yang-Mills field strength.
Integrating (12) over the $K3$, we find that there should be
$n_{1,2}$ Yang-Mills instantons in the two $E_8$s, with
\begin{equation}
n_{1} + n_{2} = 24~. 
\label{13}
\end{equation}
It is then natural to ask:  How are vacua with different choices
of $n_{1,2}$ connected to each other?

Consider the $n_1$ instantons in one $E_8$ wall.  Instanton moduli
space has singularities, including points where a single instanton
shrinks to \lq\lq zero size."  In the heterotic theory, this small
instanton can now be represented as a fivebrane sitting at the
$E_8$ wall.  But, now there is a new branch in the moduli space of
vacua - in addition to re-expanding into a large $E_8$ instanton,
the fivebrane can move off into the $S^1/Z_2$ interval!  In the process,
one loses 29 hypermultiplet moduli (the moduli of one $E_8$ instanton)
and gains a single tensor multiplet (the real scalar parametrizes the
position of the fivebrane in the interval).
Hence, one is left with $n_{1}-1$ instantons on the $E_8$ wall.
By moving across the interval and entering the other wall as an
instanton, the fivebrane can effect a transition from a vacuum 
with instanton distribution $(n_{1},n_{2})$ to a vacuum with
instanton distribution $(n_{1}-1, n_{2} + 1)$.  In this way,
the perturbative heterotic vacua with different numbers of
instantons in the two $E_8$s are all connected \cite{ganhan,sw}. 
More generally, one can modify equation (13) to read
$n_{1} + n_{2} + n_{5} = 24$, where now $n_5$ is the number
of five-branes in the interval \cite{dmw}.

We have glossed over an important point here:  The (1,0) tensor
multiplet (on the fivebrane worldvolume) contains a 
self-dual tensor $B_{\mu\nu}^{+}$.   No conventional mass
term is possible for the tensor, since there is no
$B_{\mu\nu}^{-}$ that $B_{\mu\nu}^{+}$ can pair up with. 
So, how can transitions changing the number of tensor multiplets
ever occur?

When the transition occurs, we are precisely in the situation
described in \S2.2, where there is an interacting (1,0) superconformal
field theory with $E_8$ global symmetry.  There is no weakly coupled
description of this fixed point, and a phase transition can occur there.
By going through this nontrivial fixed point, it is possible to 
connect the two branches with different numbers of tensor multiplets.
So, the novel theories of \S2 are of apparent use in unifying 6d
(0,1) supersymmetric vacua.

In fact, related theories also seem to play an important role
in connecting 4d $N=1$ vacua.  For instance, one can compactify 
the $E_8 \times E_8$ heterotic string on a Calabi-Yau threefold
$M$ which is a $K3$ fibration.
In many examples, one finds a low-energy theory with $\it chiral$
gauge representations.  For instance, if one has embedded an
$SU(3)$ bundle $V$ with $c_{1}(V) = 0$ in one of the $E_8$s, 
the unbroken subgroup of $E_8$ is $E_6$.
The matter fields come in the ${\bf 27}$ and ${\bf \overline{27}}$
representations, and one finds for the 
net number $N =  
\vert \# {\bf 27} - \# {\bf \overline{27}} \vert$ of generations: 
\begin{equation}
N = {1\over 2} \vert \int_{M} c_{3}(V) \vert~. 
\label{14}
\end{equation} 
Among the singular loci in the moduli space of vacua, there are 
places where $V$ develops a curve of singularities which  
corresponds to a small instanton in the generic $K3$ fiber of $M$.
One can represent this small instanton as a fivebrane wrapping the
base of the fibration.  In certain cases, there is a new branch
of the moduli space where the wrapped fivebrane can move away
from the $E_8$ wall into the
$S^1/Z_2$.  It was argued in \cite{kacsil} that in many cases  
this changes the net number of generations of the $E_6$ gauge theory
remaining on the wall.  
Related phenomena were discussed in \cite{aldazabal,brunner}.
So, phase transitions through close relatives
of the $E_8$ fixed point in six dimensions can also connect up 
4d string vacua with different net numbers of generations. 

\section{A Proposed M(atrix) Description}

In \S3, we have seen several interesting applications of the new
interacting 6d field theories.  However, for both the (2,0) supersymmetric
theories and the (1,0) theories with $E_8$ global symmetry, there is
no obvious definition of the theory that doesn't involve an
embedding in M-theory.  This is a very $\it un-economical$ way
of defining a quantum field theory --
one starts with far too many degrees of freedom, and must
decouple most of them from the quantum field theory of interest.

An alternative way of describing the (2,0) theories was proposed in
\cite{abkss,edsigma}, and extended to the (1,0) theories in
\cite{lowe,abks}.  We will discuss the simplest
case -- the $A_{k-1}$ (2,0) theory, i.e.
the theory of $k$ coincident M 5-branes. 
We know several suggestive facts about this theory:

\noindent$\bullet$ If we compactify
the 6d theory on a circle with radius $R$, it produces a
5d $U(k)$ Super Yang-Mills theory with coupling $g_{5}^{2} = R$.

\noindent$\bullet$ The Kaluza-Klein particles (with $p_{5} = {1\over R}$)
are \lq\lq instantons" of the $U(k)$ gauge theory (i.e., 4d Yang-Mills
instantons which look like particles in 5d).

\noindent These facts suggest that, in analogy with the M(atrix) 
approach to M-theory \cite{bfss}, we should search for a light-cone
quantization of the $A_{k-1}$ (2,0) theory.

\subsection{DLCQ of (2,0) $A_{k-1}$ Theory}

Let us take our 5+1 dimensions to be parametrized by $X^0,\cdots,X^5$.
In normal light-cone quantization, one defines $X^{\pm} = X^0 \pm X^1$
and gives initial conditions on a surface of fixed $X^+$.  Then,
one evolves forward in light-cone \lq\lq time" using the Hamiltonian
$H = P_{+}$.  The modes of quantum fields with $P_- < 0$ are
canonical conjugates of modes of $P_- > 0$, so we can choose the
vacuum to be annihilated by the $P_- < 0$ modes.  The $P_- = 0$
modes are not dynamical (but can give rise to subtleties, which will
be mentioned below).

In $\it discrete$ light-cone quantization (or DLCQ) \cite{susskind}, one 
in addition
compactifies the light-like direction
\begin{equation}
X^- \simeq X^- + 2\pi R
\label{15}
\end{equation}
Then, $P_{-}$ is quantized in units of $1/R$.
For finite $N$, the DLCQ Fock space is very simple, since there are 
a finite number of modes.
However integrating out the zero momentum modes can, 
in the DLCQ of $\it some$ theories,
lead to complicated interactions \cite{hellpol}.
The decompactification limit (where one recovers Lorentz invariance) 
is taken by going to large $R$ at fixed $P_-$, which is equivalent to
going to large $N$. 

Following \cite{seiberg}, one may find it fruitful to view the
compactification of $X^-$ as the limit of a space-like
compactification
\begin{equation}
X^- \simeq X^- + 2\pi R,~~X^+ \simeq X^+ + {R_s^{2}\over R}
\label{16}
\end{equation}
where $R_{s} \rightarrow 0$.  For finite $R_s$, one can boost
this to a spatial compactification
\begin{equation}
X^{1} \sim X^{1} + R_{s}, ~~P_{-}={N\over R} \rightarrow P_{1} = {N\over
R_{s}}
\label{17}
\end{equation}

Then, our interest is in describing modes of momentum $P_1$ as 
$R_s \rightarrow 0$.  In many cases, this can be rather complicated.
But for the (2,0) superconformal theories, we get a very weakly coupled
($g^{2} = R_s$) $U(k)$ Super Yang-Mills theory in 4+1 dimensions, with
$N$ \lq\lq instantons" (carrying charge under $J = * (F\wedge F)$). 
If we want these to have finite energy in the $\it original$
reference frame, they must have very small velocities.  So for
$R_s \rightarrow 0$, it seems that we should get a quantum mechanical
sigma model on the moduli space of $N$ $U(k)$ instantons.
The space-time supersymmetry implies that this sigma model should have
8 supercharges.

We will now give a more direct derivation of the relevant quantum
mechanics from M(atrix) theory.

\subsection{Derivation from M(atrix) Theory}

Following Berkooz and Douglas \cite{berdoug}, we know that the
background of $k$ longitudinal 5-branes 
in M(atrix) theory can be represented by
studying the theory of $N$ zero branes in the background of  
$k$ D4 branes in Type IIA string theory.  The presence of the
D4 branes (and the consequent 0-4 strings) break the supersymmetry
of the quantum mechanics to $N=8$.

The resulting quantum mechanics is in fact the dimensional reduction
of a 6d (0,1) supersymmetric system, which is a $U(N)$ gauge theory
with $k$ fundamental hypermultiplets and an additional adjoint hyper.    
The quantum mechanics has a $U(N)$ gauge symmetry and an 
$SO(4)_{\parallel} \times SO(5)_{\perp} \times U(k)$
global symmetry.
The bosonic fields are $X_\perp$, $X_\parallel$ and $q$ with
charges $\bf{(N^2,1,5,1)}$, $\bf{(N^2, (2,2),1,1)}$ and
$\bf{(N,(2,1),1,k)}$ under the gauge and global symmetries. 
Roughly speaking, $X_\perp$ characterizes the positions of
the 0 branes transverse to the 4-branes, while $X_\parallel$
characterizes their positions in the directions along the
4-branes.

The moduli space of vacua has various branches, but two are
of particular interest to us \footnote{Of course strictly
speaking the quantum mechanics has no moduli space of vacua, but
as usual in discussions of M(atrix) theory we  
can imagine a moduli space in the Born-Oppenheimer approximation.}:

\noindent 1) The Coulomb branch

On this branch, $X_\perp,
X_\parallel  \neq 0$ 
while $q = 0$.  This is the branch where the 0 branes are moving around
in spacetime away from the D4 branes, as depicted in Figure 3
below.  

\vskip .25cm
$$
\vbox{
{\centerline{\epsfxsize=3in \epsfbox{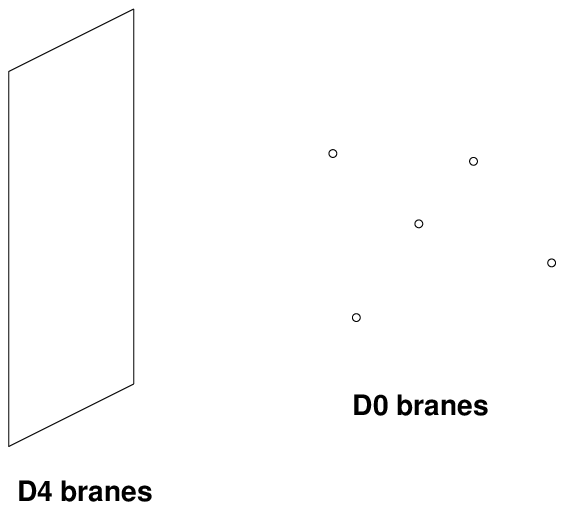}}}
{\centerline{ Figure 3:
A picture of a point on the Coulomb branch.}}
}
$$
\vskip .25cm

\noindent 2) The Higgs branch

On this branch $X_\perp = 0$ while $X_\parallel, q \neq 0$.
The D0 branes are inside of the D4 branes, as shown below
in Figure 4.
We will denote the Higgs branch moduli space for given $N$
and $k$ by ${\cal M}_{N,k}$.

So roughly speaking, the Higgs branch is concerned with the
physics on the fivebranes while the Coulomb branch captures
the physics away from the fivebranes (e.g. 11d supergravity).
This leads us to believe that the quantum mechanics
on the Higgs branch will offer a M(atrix) description of
the interacting field theories discussed in \S2.1. 
More precisely, if we want to decouple gravity from the
physics on the fivebranes, we need to take the
limit
$M_{pl} \rightarrow \infty$.  Then in particular, the coupling
in the quantum mechanics $g_{QM}$ which is related to $R$ and
$M_{pl}$ by $g_{QM}^{2} = R^3 M_{pl}^{6}$ also satisfies
$g_{QM} \rightarrow \infty$.  This is the infrared limit of
the quantum mechanics.

\vskip .25cm
$$
\vbox{
{\centerline{\epsfxsize=3in \epsfbox{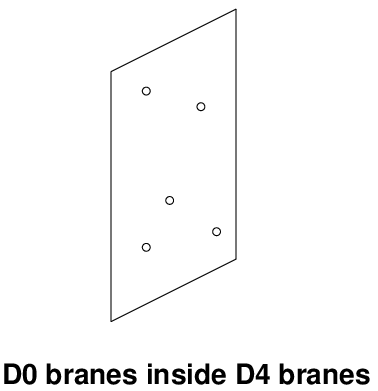}}}
{\centerline{ Figure 4:
A picture of a point on the Higgs branch.}}
}
$$ 
\vskip .25cm

The surprising fact is that in this strong coupling limit, many
simplifications occur:

\noindent 1) For $g_{QM} \rightarrow \infty$, the
Higgs branch physics
decouples from the Coulomb branch!  This is because the masses of
the massive $W$ bosons go off to infinity with $g_{QM}$, and there
is a tube of infinite length in the Coulomb branch as one 
approaches $X_{\perp} \rightarrow 0$.
So, we are left with quantum mechanics on the Higgs
branch ${\cal M}_{N,k}$.

\noindent 2) By the ADHM construction, ${\cal M}_{N,k}$ 
is actually the moduli space of $N$ instantons in $U(k)$ gauge
theory \cite{adhm}!  This is in accord with the fact that D0 branes
are expected to behave like instantons in D4 branes \cite{douglas}. 

\noindent 3) By theorems about the absence of 
couplings between vector and hypermultiplets, the Higgs 
branch does $\it not$ receive quantum corrections.  Therefore,
even the strong coupling $g_{QM} \rightarrow \infty$ limit does
not correct the sigma model on ${\cal M}_{N,k}$.

We conclude that the $A_{k-1}$ (2,0) theory has a description
in terms of the quantum mechanics on the moduli space of 
$N$ $U(k)$ instantons in the $N \rightarrow \infty$ limit.
One can similarly derive a M(atrix) description for the
$E_8$ theories with (1,0) supersymmetry \cite{lowe,abks}.
We will not review that here.

\section{Conclusions}

I have tried to emphasize in this talk 
the fruitful interaction that has occurred
in the past year or two between three different research
directions.
The new
interacting theories in $d \geq 4$ seem to have 
important uses in both the quest for a nonperturbative
definition of (compactified) M-theory, and in the search
for resolutions to the vacuum degeneracy problem. 

Several advances have been made in areas very closely
related to my talk since the conference in Buckow. 
I summarize some recent developments and future
directions here:

\noindent $\bullet$ One can try to use similar
M(atrix) formulations to study other interesting field theories,
for instance familiar 4d field theories \cite{ganseth}.

\noindent $\bullet$ One can use the quantum mechanical
formulation of the 6d theories to try and compute quantities
of interest in these conformal field theories, e.g. correlation
functions of various local operators \cite{abs}.

\noindent $\bullet$ One can try to use the
M(atrix) 
descriptions of 
the \lq\lq little string theories" of \cite{natinew} to
compute interesting properties of these novel theories without
gravity.

\noindent $\bullet$ One can 
investigate M-theory compactification
on six and higher dimensional manifolds in the M(atrix) 
formulation.  There are problems with obtaining
a simple M(atrix) description of $T^6$ compactifications (as
discussed in e.g. \cite{seiberg}),
while the situation seems to be better for Calabi-Yau
compactifications \cite{kls}. 
Successful definitions of the M(atrix) compactifications 
will involve new theories without gravity.  

\noindent $\bullet$ One can further pursue the study of
interesting phase transitions in 4d $N=1$ string vacua
\cite{kacsil,aldazabal,brunner} by using new constructions
(utilizing D-branes or F-theory) to find simple examples. 

\bigskip
\noindent
{\bf{Acknowledgements}}
\medskip
\noindent

The results discussed in this
talk were obtained in collaboration with
O. Aharony, M. Berkooz,
N. Seiberg, and E. Silverstein.  
This work was supported in part by NSF grant PHY-95-14797,
by DOE contract DOE-AC03-76SF00098, and by a DOE Outstanding
Junior Investigator Award.



\begin{thebibliography}{77}
\bibitem{sixd}
E. Witten, \lq\lq Comments on String Dynamics," hep-th/9507121.
\bibitem{st}
A. Strominger, \lq\lq Open P-branes," Phys. Lett. {\bf 383B} (1996) 44,
hep-th/9512059.
\bibitem{town} 
P. Townsend, \lq\lq D-branes from M-branes," hep-th/9512062.  
\bibitem{ganhan} 
O. Ganor and A. Hanany, \lq\lq Small $E_8$ Instantons and
Tensionless Noncritical Strings," Nucl. Phys. {\bf B474} (1996) 122,
hep-th/9602120. 
\bibitem{sw} 
N. Seiberg and E. Witten, \lq\lq Comments on String Dynamics in
Six Dimensions," Nucl. Phys. {\bf B471} (1996) 121, hep-th/9603003.
\bibitem{horwit}
P. Horava and E. Witten, \lq\lq Heterotic and Type I String
Dynamics from Eleven-Dimensions," Nucl. Phys. {\bf B460} (1996)
506, hep-th/9510209.
\bibitem{dmw}
M.J. Duff, R. Minasian, and E. Witten, \lq\lq Evidence for
Heterotic-Heterotic Duality," Nucl. Phys. {\bf B465} (1996)
413, hep-th/9601036.
\bibitem{bfss}
T. Banks, W. Fischler, S. Shenker and L. Sussind, \lq\lq M theory
as a Matrix Model: A Conjecture," Phys. Rev. {\bf D55} (1997) 112,
hep-th/9610043.
\bibitem{susskind}
L. Susskind, \lq\lq Another Conjecture About Matrix Theory," hep-th/9704080. 
\bibitem{rozali}
M. Rozali, \lq\lq Matrix Theory and U Duality in Seven Dimensions,"
Phys. Lett. {\bf 400B} (1997) 260, hep-th/9702136.
\bibitem{brs}
M. Berkooz, M. Rozali and N. Seiberg, \lq\lq Matrix Description of
M theory on $T^4$ and $T^5$," Phys. Lett. {\bf 408B} (1997) 105, 
hep-th/9704089.
\bibitem{kacsil}
S. Kachru and E. Silverstein, \lq\lq Chirality Changing Phase
Transitions in 4d String Vacua,'' Nucl. Phys. {\bf B504} (1997)
272, hep-th/9704185. 
\bibitem{aldazabal}
G. Aldazabal, A. Font, L. Ibanez, A. Uranga and G. Violero, \lq\lq 
Non-Perturbative Heterotic D=6, D=4, N=1 Orbifold Vacua,"
hep-th/9706158.
\bibitem{brunner}
I. Brunner, A. Hanany, A. Karch and D. L\"ust, \lq\lq Brane Dynamics
and Chiral Nonchiral Transitions," hep-th/9801017.
\bibitem{abkss}
O. Aharony, M. Berkooz, S. Kachru, N. Seiberg and E. Silverstein,
\lq\lq Matrix Description of Interacting Theories in Six Dimensions,"
hep-th/9707079.
\bibitem{edsigma}
E. Witten, \lq\lq On the Conformal Field Theory of the Higgs Branch,"
hep-th/9707093.
\bibitem{lowe}
D. Lowe, \lq\lq $E_8 \times E_8$ Small Instantons in Matrix Theory,"
hep-th/9709015. 
\bibitem{abks}
O. Aharony, M. Berkooz, S. Kachru and E. Silverstein, \lq\lq Matrix
Description of (1,0) Theories in Six Dimensions," hep-th/9709118.
\bibitem{hellpol}
S. Hellerman and J. Polchinski, \lq\lq Compactification in the
Lightlike Limit," hep-th/9711037. 
\bibitem{seiberg}
N. Seiberg, \lq\lq Why is the Matrix Model Correct," 
Phys. Rev. Lett. {\bf 79} (1997) 3577, hep-th/9710009.
\bibitem{berdoug}
M. Berkooz and M. Douglas, \lq\lq Five-branes in M(atrix) Theory,"
Phys. Lett. {\bf 395B} (1997) 196, hep-th/9610236. 
\bibitem{adhm}
M. Atiyah, V. Drinfeld, N. Hitchin and Y. Manin, \lq\lq Construction
of Instantons," Phys. Lett. {\bf 65B} (1978) 185.
\bibitem{douglas}
M. Douglas,\lq\lq Branes within Branes," hep-th/9512077.
\bibitem{ganseth}
O. Ganor and S. Sethi, \lq\lq New Perspectives on Yang-Mills Theories
with Sixteen Supersymmetries," hep-th/9512071. 
\bibitem{abs}
O. Aharony, M. Berkooz and N. Seiberg, \lq\lq Light-Cone Description
of (2,0) Superconformal Theories in Six Dimensions," hep-th/9712117.
\bibitem{natinew}
N. Seiberg, \lq\lq New Theories in Six Dimensions and Matrix Description
of M Theory on $T^5$ and $T^5/Z_2$," Phys. Lett. {\bf 408B} (1997) 98,
hep-th/9705221. 
\bibitem{kls}
S. Kachru, A. Lawrence and E. Silverstein, \lq\lq On the Matrix
Description of Calabi-Yau Compactifications," hep-th/9712223.




\end{thebibliography}
\end{document}